\documentclass{WileyMSP-template}

\usepackage{bm}
\usepackage{authblk}
\usepackage{amsmath}
\usepackage{graphicx}
\usepackage[section]{placeins}
\usepackage{xcolor}
\usepackage{wrapfig}
\usepackage{wrapfig,lipsum,booktabs}
\usepackage{tabularx}
\usepackage{float}
\usepackage{gensymb}
\usepackage[utf8]{inputenc}
\makeatletter
\let\saved@includegraphics\includegraphics
\AtBeginDocument{\let\includegraphics\saved@includegraphics}
\renewenvironment*{figure}{\@float{figure}}{\end@float}
\makeatother
\usepackage[labelfont=bf]{caption}
\usepackage{cite}
\usepackage{soul}

\usepackage{hyperref}
\hypersetup{colorlinks=true, linkcolor=blue, citecolor=blue, urlcolor=blue,}
\usepackage{soul}
\usepackage{subcaption}
\usepackage{cleveref}

\usepackage{setspace}

\begin{document}



\RaggedRight 
\setlength{\parindent}{2em}
\RaggedRight \title{Mechanical Control of Polar Order}

\maketitle


\author{Pushpendra Gupta$^\dagger$,$^\ast$}
\author{Peter Meisenheimer$^\dagger$}
\author{Xinyan Li}
\author{Sajid Husain}
\author{Vishantak Srikrishna}
\author{Ashley Cortesis}
\author{Yimo Han}
\author{Ramamoorthy Ramesh a,c,d$^\ast$}

\begin{affiliations}

\RaggedRight Pushpendra Gupta$^\ast$\footnotemark[1]\\ 
Department of Materials Science and Engineering, University of California, Berkeley, CA 94720, USA\\
Email Address: pushpendra@berkeley.edu\\

Peter Meisenheimer\footnotemark[1]\\
Kepler Computing, 617 River Oaks Pkwy, San Jose, CA, 95134, USA\\
Email Address: peter@keplercompute.com\\

Xinyan Li\\
Rice Advanced Materials Institute, Rice University, Houston, TX 77005, USA.\\
Email Address: xl156@rice.edu\\

Sajid Husain\\
Department of Materials Science and Engineering, University of California, Berkeley, CA 94720, USA\\
Email Address: shusain@berkeley.edu\\

Vishantak Srikrishna\\
Kepler Computing, 617 River Oaks Pkwy, San Jose, CA, 95134, USA\\
Email Address: vishantak@keplercompute.com\\

Ashley Cortesis\\
Department of Engineering Physics, University of California, Berkeley, CA 94720, USA\\
Email Address: acortessis@berkeley.edu\\

Yimo Han\\
Rice Advanced Materials Institute, Rice University, Houston, TX 77005, USA.\\
Email Address: yh76@rice.edu\\

Ramamoorthy Ramesh$^\ast$\\
Department of Materials Science and Engineering, University of California, Berkeley, CA 94720, USA\\
Rice Advanced Materials Institute, Rice University, Houston, TX 77005, USA.\\
Department of Physics, University of California, Berkeley, CA 94720, USA\\
Email Address: rramesh@berkeley.edu\\

\footnotetext[1]{These authors contributed equally.}

\end{affiliations}



\begin{abstract}

BiFeO$_{3}$ is a model multiferroic in which the ferroelectric polarization is coupled to ferroelastic lattice distortions, yet deterministic control of its domain structure remains limited by high switching fields and competing polarization variants. Here we identify a mechanically assisted polarization switching pathway in epitaxial BiFeO$_{3}$ thin films that fundamentally alters the switching energetics. Using just out-of-plane electric fields, polarization reversal requires voltages of approximately 4 V and stabilizes coexisting polarization states. In contrast, when mechanical pressure is applied concurrently, the coercive voltage can be significantly reduced, (even to 0V), resulting in spontaneous switching. Piezoresponse force microscopy measurements reveal that applied mechanical pressure suppresses ferroelastic domain competition, indicating a decrease in the required electrical energy barrier associated with polarization rotation and domain-wall motion. These results demonstrate that stress acts as an active thermodynamic control parameter, enabling access to switching pathways that are inaccessible under only an electric field. By directly coupling lattice distortions to polarization reversal, mechanically assisted switching provides a general framework for controlling coupled order parameters in multiferroic oxides, which can be directly applied in the device-level architecture, where a small mechanical pressure can help in achieving lower switching energy of ferroelectric polarization. This work advances the fundamental understanding of electromechanical coupling in complex ferroics and establishes mechanical energy as a powerful tool for probing and manipulating ferroelastic–ferroelectric interactions.

\end{abstract}


\vspace{1 cm} 
\section{Introduction}
The controlled manipulation of order parameters by external stimuli is a central problem in condensed-matter physics, as it provides a means to probe and exploit collective phenomena governed by symmetry, lattice distortions, and long-range interactions \cite{wang2025manipulation}. In multiferroic materials, where multiple ferroic orders coexist and are coupled, the response to external fields is inherently complex, since driving one order parameter inevitably perturbs others through elastic and electrostatic coupling mechanisms \cite{spaldin2005renaissance,eerenstein2006multiferroic,ramesh2007multiferroics}. Understanding how these coupling influence switching pathways and domain evolution remains a key challenge, particularly in systems in which ferroelastic distortions play a dominant role.

BiFeO$_3$ (BFO) is a prototypical room-temperature multiferroic and has been widely studied as a model system for coupled ferroelectric and ferroelastic behavior \cite{wang2003epitaxial,catalan2009physics}. It crystallizes in a rhombohedrally distorted perovskite structure (space group \textit{R3c}), characterized by a large spontaneous ferroelectric polarization along the pseudocubic $\langle 111 \rangle$ directions, antiphase rotations of the oxygen octahedra, and G-type antiferromagnetic order \cite{sosnowska1982spiral,kubel1990structure}. Importantly, the polarization, lattice strain, and octahedral rotation modes in BFO are inter-dependent; symmetry-allowed coupling terms in the free energy link these order parameters, giving rise to multiple ferroelectric and ferroelastic variants that are closely spaced in energy \cite{Heron2014-gu}.

In epitaxial thin films, substrate-induced strain and reduced symmetry further modify this energy landscape, stabilizing specific domain variants and domain-wall configurations. As a result, polarization switching in BFO thin films is generally not a simple 180$^\circ$ reversal, but involves ferroelastic domain wall motion and polarization rotation through intermediate crystallographic states\cite{Heron2014-gu}. Experimental studies using piezoresponse force microscopy (PFM) and related techniques have shown that electric-field-driven switching frequently produces mixed ferroelastic domain configurations, even under relatively large applied fields \cite{balke2012enhanced}. These observations indicate that the energetic cost of lattice reorientation and strain accommodation plays a decisive role in limiting deterministic domain control.

From a thermodynamic standpoint, the coupled behavior of polarization and strain in BFO can be described within Landau--Ginzburg--Devonshire-type formalisms that include electromechanical coupling between polarization and elastic degrees of freedom \cite{tagantsev2013origin}. Such models predict that polarization switching pathways and energy barriers depend sensitively on elastic boundary conditions and on the relative stability of competing ferroelastic variants. Consequently, purely electric-field-driven switching may be insufficient to fully overcome the energetic barriers associated with ferroelastic reconfiguration, leading to incomplete switching or the stabilization of multidomain states.

Mechanical stress provides a complementary route to influence ferroic systems by coupling directly to lattice distortions. In ferroelectric and ferroelastic materials, applied stress/strain is known to modify domain stability and domain-wall motion by altering the elastic contribution to the free energy \cite{damjanovic1998ferroelectric}. In multiferroics such as BFO, theoretical work has suggested that mechanical boundary conditions can significantly affect polarization rotation pathways by selectively stabilizing specific ferroelastic variants \cite{tagantsev2013origin}. Experimentally, static strain engineering through epitaxial growth has been successfully employed to access strain-stabilized phases and enhanced functional responses in multiferroics/ferroelectrics \cite{zeches2009strain}. However, epitaxial strain is static and does not provide dynamic control over ferroelastic domain evolution during polarization switching.

In this work, we investigate polarization switching in epitaxial BFO thin films under the combined influence of electric fields and localized mechanical pressure \cite{gruverman2019piezoresponse}. We show that switching via electric field requires relatively high voltages and results in the persistence of multiple polarization variants, consistent with previous reports\cite{wang2003epitaxial}. When mechanical pressure is concurrently applied, however, the switching voltage is reduced significantly and the domain configuration evolves toward a single majority domain state. The mechanical stress modifies the elastic contribution to the coupled ferroelectric-ferroelastic energy landscape by giving direct energy to the ferroelastic domain and lowering the energy barriers associated with lattice deformation. These results highlight the importance of lattice-mediated effects in determining polarization switching pathways and establish mechanically assisted switching as an effective approach for controlling coupled order parameters in multiferroic oxides.

\section*{Results and Discussion}

\begin{figure}[t!]
\centering
\includegraphics[width=\linewidth]{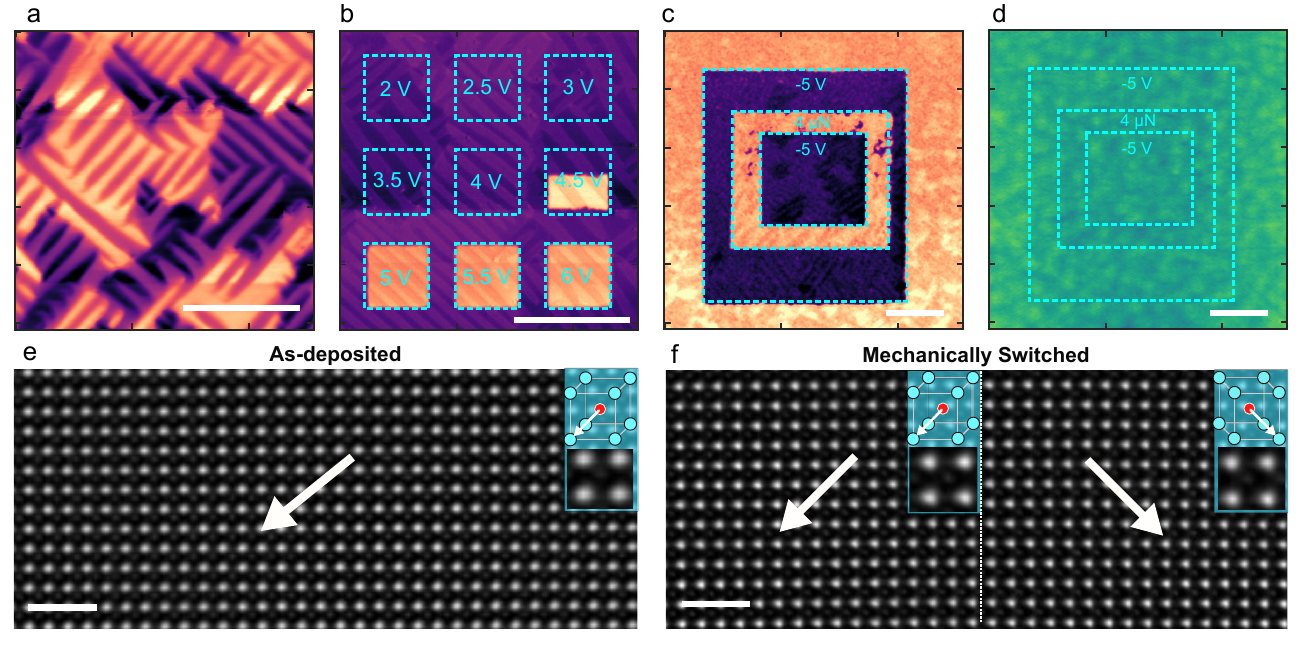}
\caption{Voltage-assisted and mechanically-assisted polarization switching in BiFeO$_3$ thin films. (\textit{A}) In-plane piezoresponse image of an as-grown BiFeO$_3$/SrRuO$_3$ thin film on a SrTiO$_3$ (001) substrate, showing the canonical 4-variant ferroelectric domains. 
(\textit{B}) Out-of-plane piezoresponse image acquired after the application of DC voltage, demonstrating that a bias of approximately 4~V is required to induce polarization switching under electric field. 
(\textit{C}) Box-in-box (BiB) piezoresponse image recorded following sequential application of voltage, mechanical force, and voltage. Application of a force of 4~$\mu$N induces clear domain switching, and the mechanically switched domains can be reversibly re-switched using an electric field. 
(\textit{D}) Corresponding surface topography of the marked region, confirming the absence of measurable surface damage after force-assisted switching. Scale bars are 2 $\mu$m.
(\textit{E-f}) High-angle annular dark-field scanning transmission electron microscopy (HAADF-STEM) images acquired from the as-deposited region and the mechanically switched region. Comparison of these images reveals no observable structural defects or lattice disruption associated with mechanical switching. The insets are the zoomed images and schematics of the unit cell. Scale bars are 1 nm.}
\label{fig:Fig1}
\end{figure}

Epitaxial BiFeO$_3$ (BFO) thin films (65~nm) were grown on SrTiO$_3$(001) substrates with a 10~nm SrRuO$_3$ bottom electrode by pulsed laser deposition. The deposition was performed at a substrate temperature of $720^{\circ}\mathrm{C}$ under an oxygen pressure of $100\,\mathrm{mTorr}$ and $150\,\mathrm{mTorr}$ for SRO and BFO, respectively. The laser energy density was maintained $0.8\,\mathrm{J/cm}^2$ with a repetition rate of $10\,\mathrm{Hz}$ for SRO and $1.2\,\mathrm{J/cm}^2$ with a repetition rate of $4\,\mathrm{Hz}$ for BFO. The film quality is confirmed with X-ray diffraction and atomic force microscopy (AFM), shown in Supp. Fig. 1. Piezoresponse force microscopy (PFM) of films shows the canonical, 4-variant ferroelectric domain pattern that is expected for BFO on SrTiO$_3$, shown in \textbf{Figure \ref{fig:Fig1}a}. In this work, we perform a detailed analysis of the electrically and mechanically driven switching of BFO thin films (Methods).

\subsection*{Out-of-plane polarization switching}
To first demonstrate the switchable ferroelectric polarization in the unperturbed (positive) state, samples are first poled in a large area with a voltage of approximately -5 V, then smaller areas are switched back to the positive orientation using variable voltages (\textbf{Figure \ref{fig:Fig1}b}. Here, polarization reversal is observed only above +4~V and this defines the effective coercive voltage ($V_C$). Strikingly, we find that in lieu of the positive voltage, pressure (approximately 4 $\mu$N) from the AFM tip alone can switch the sample back into the positively poled configuration (\textbf{Figure \ref{fig:Fig1}c, Supp. Figure 2}). Although this effect appears to be unidirectional, i.e., it can only switch the ferroelectric from the negative to the positive state, it can be fully reversed by a successive negative voltage pulse and then subsequently switched again using mechanical force. An example of this repeated cycling is shown in \textbf{Figure \ref{fig:Fig1}c} and when this mechanically switched area is investigated topographically, the atomic steps are identical to those of pristine, indicating that there is no irreversible mechanical deformation of the film. This scheme then shows a distinct, purely mechanical method of switching and writing ferroelectric domains.

\begin{figure}[t!]
\centering
\includegraphics[width=1\linewidth]{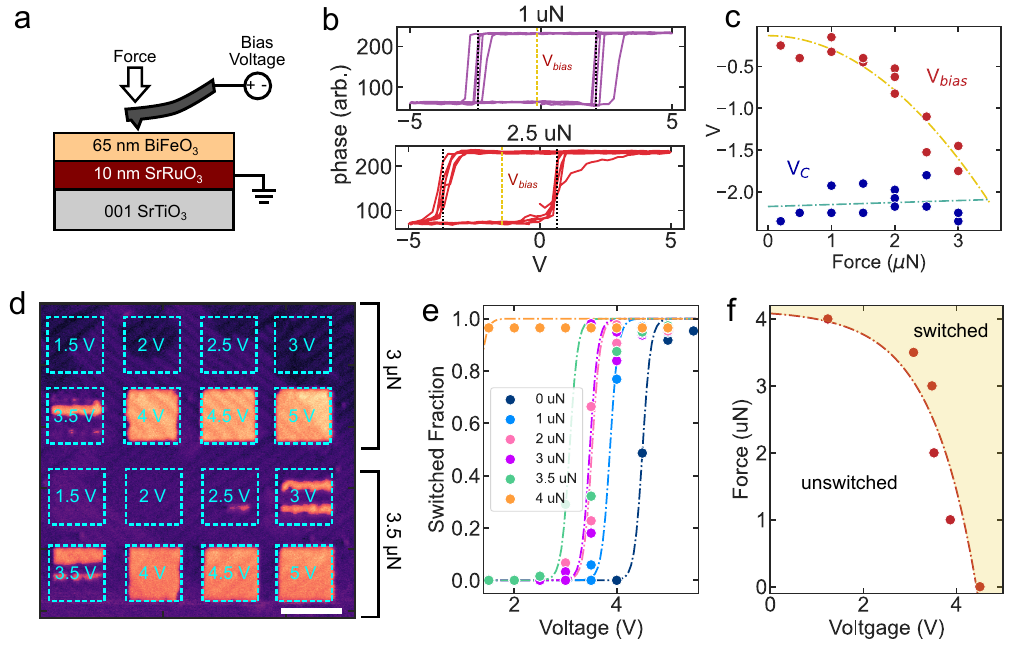}
\caption{\textbf{Mechanically-assisted switching of polarization in BiFeO$_3$ thin films.}
(\textit{A}) Schematic illustration of the experimental configuration for applying electric bias and mechanical force using a conductive AFM tip. The epitaxial SrRuO$_3$ bottom electrode provides an electrical ground. 
(\textit{B}) Example piezoresponse hysteresis loops acquired under varying applied mechanical forces. At low applied force, the hysteresis loop remains symmetric about the vertical axis, indicating predominantly voltage-driven polarization switching. With increasing mechanical force, the hysteresis loop progressively shifts along the voltage axis toward the negative direction. 
(\textit{C}) Coercive voltage and bias voltage extracted from the hysteresis loops as a function of applied mechanical force. While the coercive voltage remains approximately constant, the negative bias voltage increases approximately with mechanical force, indicating enhanced switching asymmetry under mechanical loading.
(\textit{D}) Example piezoresponse image recorded after mechanically-assisted switching, where the applied voltage is varied from 1.5~V to 5~V at fixed mechanical forces of 3~$\mu$N (top two rows) and 3.5~$\mu$N (bottom two rows). Increasing the mechanical force enables polarization switching at progressively lower applied voltages.
(\textit{E}) Switching fraction of polarization domains as a function of applied voltage for different mechanical forces. At a force of 4~$\mu$N, complete switching is achieved even in the absence of an external electric bias. At lower mechanical forces, a finite voltage is required to achieve deterministic switching. Measured switching fractions are fit to $tanh(V)$.
(\textit{F}) Switching voltage extracted from the fit in \textit{D} plotted with the corresponding force. Complete switching is obtained at 4~$\mu$N without applied voltage, whereas reducing the mechanical force necessitates increasing electric bias, with voltages of approximately 4.5~V sufficient to fully switch the domains in the absence of mechanical assistance.}
\label{fig:Fig2}
\end{figure}

To further verify our approach of electrical and mechanical switching, high-angle annular dark-field scanning transmission electron microscopy (HAADF-STEM) was conducted on regions corresponding to the as-deposited, electrically switched, and mechanically switched states (\textbf{Figure \ref{fig:Fig1}d-f}). In all three regions, it switches fully, the atomic lattice remains well-ordered, with no evidence of extended defects, amorphization, dislocation formation, or interface degradation within the spatial resolution of the measurements. The preservation of sharp atomic columns indicates that both electrical and force-assisted switching processes were executed without inducing detectable structural deformation. The polarization orientation in each region is inferred from the relative displacement of cation sublattices, as indicated by the arrows, and is consistent with the PFM-derived switching behavior. These observations confirm that the mechanically assisted switching is non-destructive and originates from reversible electromechanical coupling rather than irreversible lattice deformation. To then investigate the underlying physics of this interaction, we turn to combinatorial experiments where a force and a voltage are applied simultaneously, schematically shown in \textbf{Figure \ref{fig:Fig2}a}. Details of the poling scheme are shown in \textbf{Supp. Figure 5}.

First, by holding the conductive PFM cantilever at a single point and cycling the voltage, the ferroelectric loop can be effectively traced at a single point. We note that, as the voltages in the AFM are subject to a number of factors such as tip contact area or geometrically enhanced electric fields, the voltages in these measurements may not be quantitatively comparable to those in other experiments. When the normal force of the cantilever is low, around 1 $\mu$N, the piezoelectric loop is approximately centered with respect to voltage. As the downward force of the cantilever is systematically increased, the piezoelectric loop is proportionally shifted to negative voltage, indicating that the force acts as an effective positive voltage. At approximately 3 $\mu$N, the loop is shifted such that the positive coercive voltage is almost 0, indicating that it is approaching spontaneous positive switching. Example piezoelectric loops are shown in \textbf{Figure \ref{fig:Fig2}b}. The bias voltage ($V_{bias}$) and coercive voltage ($V_C$) extracted from these measurements are plotted in \textbf{Figure \ref{fig:Fig2}c}. From these data, we observe that the bias voltage shifts negatively with applied force, where the coercive voltage remains approximately constant. This indicates that mechanical force is acting as an effective voltage, shifting the energy landscape asymmetrically. Additionally, extrapolation of these voltages indicates a crossing point of around 4 $\mu$N where the force results in spontaneous positive switching, which agrees exactly with the force used for mechanical switching in \textbf{Figure \ref{fig:Fig1}c}. These results are robust when cycling the force from 1 $\mu$N to 3 $\mu$N repeatedly without hysteresis (\textbf{Supp. Figure 4}).

We then conducted a similar measurement as in \textbf{Figure \ref{fig:Fig1}b}, where a voltage is used to switch a finite area, but including mechanical assistance for positive switching. Example PFM images are shown in \textbf{Figure \ref{fig:Fig2}d}, where out-of-plane switching occurs at lower voltages when the force of the PFM tip is increased. These images can then be analyzed to determine an out-of-plane switching fraction as a function of both voltage and force. These extracted data are plotted in \textbf{Figure \ref{fig:Fig2}e}, where the switched fraction per voltage is fit to $tanh(V-V_C')$. Comparing to \textbf{Figure \ref{fig:Fig2}c}, $V_C'$ would map to the difference $V_{bias}-V_C$. Immediately, it is evident that electrical switching becomes easier with increasing force, as expected from the results discussed above. $V_C'$ extracted from these fits is then plotted as a function of voltage and force in \textbf{Figure \ref{fig:Fig2}f}, which can be used to map an effective phase diagram of ferroelectric switching. This again shows that mechanical force can be used to aid electrical switching up to a point, after which only mechanical force is required for spontaneous ferroelectric switching.

\begin{figure}
\centering
\includegraphics[width=1\linewidth]{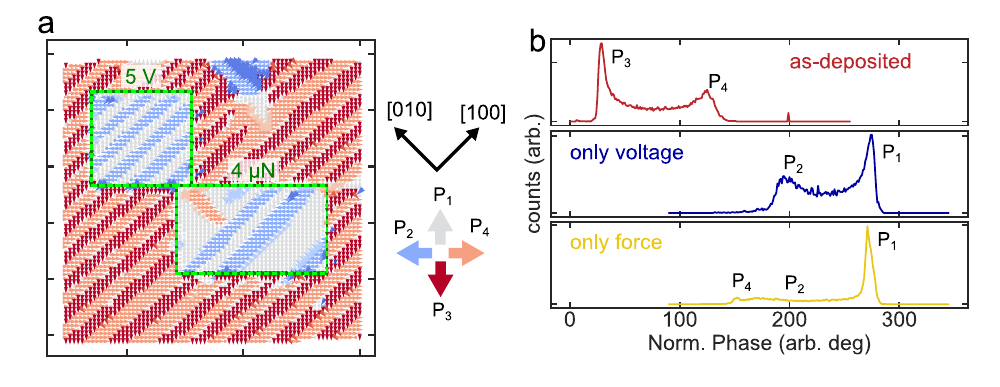}
\caption{\textbf{Voltage- and mechanically induced domain switching in BiFeO$_3$ thin films.}
(\textit{A}) Vector mapping of the polarization switching induced by electric bias and mechanical force. The outer region corresponds to switching performed using a $-5$~V bias (red and pink arrows), whereas the region enclosed by the green dotted lines indicates areas switched using a $+5$~V bias and a normal force of 4~$\mu$N, respectively. Switching induced by electric field alone results in predominantly 180$^\circ$ polarization reversal, while mechanically assisted switching produces a ferroelastic reconfiguration in which one polarization variant becomes dominant over the other. (\textit{B}) Polarization histograms extracted from the three regions. In the as-grown state, both polarization variants are present with comparable intensity. Following electric-field-only switching, the polarization undergoes 180$^\circ$ reversal while maintaining a similar proportion of domain variants. In contrast, mechanically assisted switching suppresses one variant and stabilizes the other, driving the system toward a single domain population.}
\label{fig:Fig3}
\end{figure}

\subsection*{In-plane polarization switching}
While out-of-plane ferroelectric switching is directly aided by the application of mechanical force, the in-plane orientation of the ferroelectric domains is more complex and may provide more insight into the underlying mechanism. To observe and monitor the in-plane orientation of ferroelectric domains, the 3D polarization vector can be reconstructed from one out-of-plane and two orthogonal in-plane PFM scans. These images are taken by physically rotating the sample between scans, and the vector map is reconstructed using \textit{opencv} implemented in Python to align the images (\textbf{Supp. Figure 8}). Shown in \textbf{Figure \ref{fig:Fig3}a} is a vector map with three distinct regions. The majority of the area has been poled down from the as-deposited state with -5 V and the chosen region shows primarily 2-variant 109$^\circ$ ferroelectric domains. The second region, in the top left, is poled using +5 V and minimal force. As can be seen, the domain structure is preserved in this region and individual domains are switched 180$^\circ$ in-plane. The last region is switched using only a mechanical force of approximately 4 $\mu$N, but it appears that the domain structure is not preserved. These switching pathways are quantified in \textbf{Figure \ref{fig:Fig3}b}, which shows histograms of the polarizations in the different regions. From the as-deposited to voltage switched regions, the in-plane polarizations are inverted from $P_3 + P_4$ to $P_1 + P_2$, respectively, in equal proportion. In the mechanically switched region, however, $P_4$ domains are not directly switched but are instead suppressed. As the switching pathway of domains in BFO is governed primarily by ferroelastic interactions at domain walls, preference for a single domain variant implies an elastic anisotropy and may indicate that the mechanism behind the mechanical switching is likely to be primarily flexoelectric.

\begin{figure}
\centering
\includegraphics[width=1\linewidth]{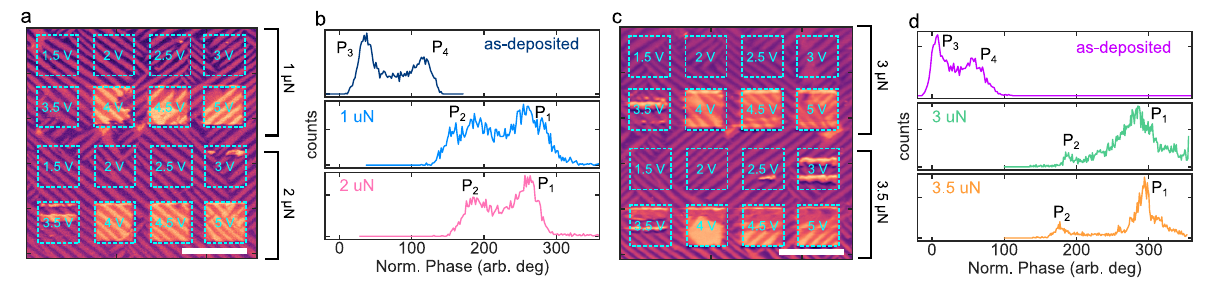}
\caption{\textbf{Mechanical switching and reconfiguration of in-plane polarization domains in BiFeO$_3$ thin films.}
(\textbf{A,C}) In-plane PFM phase images recorded after mechanical switching, with the applied voltage varied from 1.5~V to 5~V at fixed normal mechanical forces as indicated on the right. Increasing mechanical force enables polarization switching at progressively lower applied voltages and modifies the resulting domain configuration.
(\textit{B,D}) Corresponding intensity profiles extracted from the marked regions in (\textit{A,C}). In the as-grown state, both in-plane polarization variants are present with comparable intensity. At low applied mechanical force, the polarization undergoes predominantly 180$^{\circ}$ ferroelectric reversal while preserving a similar population of domain variants. In contrast, increasing mechanical force progressively suppresses one ferroelastic variant and stabilizes the other, driving the system toward a configuration with a single in-plane domain.}
\label{fig:Fig4}
\end{figure}

As above, this experiment has been repeated using a combination of forces and voltages, ultimately showing that this domain preference emerges only at relatively large forces. In \textbf{Figure \ref{fig:Fig4}a}, at forces of 1-2 $\mu$N, switching happens prior to the nominal $V_C$ (4 V) in combination with mechanical force, but the reorientation of ferroelectric domains largely preserves the structure of the unswitched state. This can be seen quantitatively in \textbf{Figure \ref{fig:Fig4}b}, where $P_3$ and $P_4$ domains reverse to $P_1$ and $P_2$, preserving their relative proportion. In \textbf{Figure \ref{fig:Fig4}c}, however, at forces $>$ 3 $\mu$N, a similar effect is seen as in \textbf{Figure \ref{fig:Fig3}b}, where $P_4$ domains are out-competed during switching and the final state prefers a majority polarization.

\subsection*{Mechanism of switching}
As the mechanical pressure of the tip appears to behave directly like an additional voltage, we consider whether this mechanism is triboelectric or flexoelectric in origin. While both techniques have been shown to create an effective voltage originating from the pressure of an AFM tip on a surface\cite{Zhou2013-jc, Lu2012-hn, Cho2024-lp, Stefani2021-hh}, there are signatures of triboelectricity which we do not observe, leading us to conclude that the effect is primarily flexoelectric. Namely, a key signature of triboelectricity is that the buildup of voltage with AFM tip pressure should be dependent on repeated contact and distancing of the tip, which is not observed here.

 The triboelectric effect is caused by mechanical friction between two dissimilar materials and results in a direct transfer of charges from one surface to the other \cite{Harper1970-hu}. This effect has been empirically shown to be a function of several parameters, including the work function difference between the materials and contact force. As the difference in electrical properties of the materials gives the sign and magnitude of the accumulated charge, our expectation is that it should be positive charges from a high work function metal on an oxide surface \cite{Harper1957-ug}. While the use of a platinum-coated AFM tip has been shown to accumulate charges up to a 1 V equivalent on a silicon oxide surface \cite{Zhou2013-jc}, there are other requirements of the effect that may not agree with the data presented here. First, the triboelectric effect is dependent on the repeated contact and removal of the two materials. While this may be true for the experiment in \textbf{Figure \ref{fig:Fig2}c}, where the tip is rastered over the surface, in the piezoelectric loops of \textbf{Figure \ref{fig:Fig2}a}, the tip does not lose contact with the surface while the force is cycled. As the charges in the triboelectric effect are transferred on first contact, this implies that the effect should not be cyclable unless the contact is reestablished with each experiment. Second, triboelectric charge transfer should be a function of the number of contacts, as the time scale for charge dissipation in an insulator is much longer than the experiment \cite{Zhou2013-jc}. In this case, one would expect the force required to spontaneously switch the polarization to be a function of cycle number, i.e., a lower force could be used for switching if rastered many times over the same region. In this case, switching appears to be thresholded near 4 $\mu$N where, even at 3.5 $\mu$N, 10 cycles does not drive any kind of partial switching (\textbf{Supp. Figure 7}).

The flexoelectric effect, in contrast, is likely to be the cause here due to an elastic gradient in the film caused by the local pressure of the AFM tip \cite{Jia2025-il, Liang2024-uh}. More generally, flexoelectricity has been used in epitaxial thin films to modify ferroelectric energy landscapes\cite{Zou2021-on}. In the flexoelectric case, the strain field from the tip would directly modify and asymmetrize the energy landscape of the ferroelectric as a function of force \cite{Lu2012-hn, Cao2017-nz, Gu2015-uq}. Additionally, reports of a flexoelectric field from an AFM tip have been reported previously \cite{Lu2012-hn, Park2018-ow} in other systems, and the expected field from even small forces is large, on the order of MV/cm. Indeed, previous experiments with other systems have reported flexoelectric coefficients of $\approx$ 1 $\mu$N V$^{-1}$, which is in excellent agreement with the values seen here. This work, however, is the first to report the flexoelectric effect as a direct aid to an applied voltage in a combinatorial switching.

If we assert that the electrical energy consumed by ferroelectric switching is $\approx2P_rV/t$, the application of mechanical force directly reduces this number though reducing the effective $V$. While the total energy consumed must be conserved, this work provides an avenue to substitute the required electric field for a mechanical contribution. Beyond purely mechanical switching of the ferroelectric, we demonstrate a method that may be used to effectively threshold the polarization switching at a smaller and constant voltage, providing the additional functionality of a mechanically-based selector or energy harvester.

\section{Conclusion}
Controlling ferroelastic and ferroelectric domains in BiFeO$_{3}$ thin films is essential to achieving their functional potential, but high electric-field needs and persistent multidomain states continue to impede practical switching. Here, we demonstrate how the switching environment is substantially changed when mechanical pressure is coupled with an out-of-plane electric field. Mechanically aided switching minimizes switching voltage and leads to a preference for a single domain population, while electrical switching requires $\approx$4 V and stabilizes coexisting polarization variations. For the first time, it is shown that these fields can be applied in combination to reduce the switching voltage, paving the way for Micro-Electro-Mechanical System (MEMS) or Nano-Electro-Mechanical Systems (NEMS) devices that can act as a ferroelectric selector memory or efficient energy harvester.  Our results open new possibilities for strain-enabled, energy-efficient ferroelectric devices and demonstrate mechanically aided switching as a reliable approach for domain engineering in multiferroic oxides.


\section{Experimental Section}
\noindent\textbf{Sample fabrication}\\
Epitaxial BiFeO$_3$ thin films with a thickness of 65~nm were grown on single-crystal SrTiO$_3$(001) substrates using a 10~nm SrRuO$_3$ bottom electrode by pulsed laser deposition (PLD). Prior to deposition, the SrTiO$_3$ substrates were chemically and thermally treated to obtain atomically flat, TiO$_2$-terminated surfaces. The SrRuO$_3$ layer was deposited at a substrate temperature of 720~$^\circ$C under an oxygen pressure of 100~mTorr, using a laser repetition rate of 10~Hz and a laser fluence of approximately 1~J\,cm$^{-2}$. Subsequently, BiFeO$_3$ films were deposited \emph{in situ} at the same substrate temperature of 720~$^\circ$C under an oxygen pressure of 150~mTorr, with a laser repetition rate of 4~Hz and a fluence of 1.2~J\,cm$^{-2}$.\\

\noindent\textbf{X-ray Diffraction}\\
  The crystalline structure and phase purity of the films were examined by high-resolution X-ray diffraction (XRD) using Cu K$\alpha$ radiation. The $\theta$--$2\theta$ scans exhibit only the (001), (002), and (003) reflections of BiFeO$_3$, confirming single-phase, $c$-axis-oriented epitaxial growth without detectable impurity phases. The SrRuO$_3$(001) reflection appears as a shoulder adjacent to the SrTiO$_3$ substrate peak, consistent with coherent epitaxial growth and the small lattice mismatch between SrRuO$_3$ and SrTiO$_3$. \\

\noindent\textbf{Atomic Force and Piezoresponse Force Microscopy}\\ 
    Surface morphology was characterized by atomic force microscopy (AFM), which reveals well-defined terrace–step features that replicate the underlying substrate morphology, indicating smooth surfaces and step-flow or layer-by-layer growth. Piezoelectric  properties were investigated by piezoresponse force microscopy (PFM) on the same instrument, employing BudgetSensors Platinum/chromium-coated probes (Multi75G; spring constant $\sim$3\,N/m, resonance frequency 75\,kHz, tip radius 25\,nm). Dual AC resonance-tracking (DART) mode was used for high-sensitivity imaging of piezoelectric responses, with drive amplitudes varied between 1\,V and 1.5\,V. For lateral signal acquisition, the probe was tuned to the corresponding contact resonance frequency of $\sim$670\,kHz. Piezoresponse images are calculated using both the amplitude, $A$, and phase, $\phi$, of the response through $A\cos{\phi}$. Vector maps are reconstructed using one out-of-plane and two, orthogonal in-plane scans to measure the $P_z$, $P_y$, and $P_x$ components respectively. The images are then overlain programmatically using the \textit{opencv} module in python.  \\

 \noindent\textbf{Scanning transmission electron microscopy}\\
 Cross-sectional samples for STEM imaging were prepared using an FEI Helios 660 focused ion beam (FIB). During the ion milling process, the acceleration voltages were gradually reduced from 30 to 5 kV and finally to 2 kV to minimize amorphous layers on the surface. HAADF-STEM images were captured using a 300 kV spherical aberration-corrected transmission electron microscope (FEI Titan Themis3 S/TEM) with a convergence angle of 25 mrad.\\

\medskip
\noindent\textbf{Supporting Information} \\ 
Supporting Information is available from the Wiley Online Library or from the author.\\

\medskip
\noindent\textbf{Acknowledgements} \\
P.G. (Pushpendra Gupta) and R.R. acknowledge the U.S. Department of Energy for support of this research work. S. H., and R.R., acknowledge that this research was sponsored by the Army Research Laboratory and was accomplished under Cooperative Agreement Number W911NF-24-2-0100. The views and conclusions contained in this document are those of the authors and should not be interpreted as representing the official policies, either expressed or implied, of the Army Research Laboratory or the U.S. Government. The U.S. Government is authorized to reproduce and distribute reprints for government purposes, notwithstanding any copyright notation herein. X.L. acknowledges support from the Rice Advanced Materials Institute (RAMI) at Rice University as a RAMI Postdoctoral Fellow. X.L. and Y.H. acknowledge support from NSF (FUSE-2329111 and CMMI-2239545) and Welch Foundation (C-2065). X.L. and Y.H. acknowledge the Electron Microscopy Center, Rice University.\\

\medskip

%
\bibliographystyle{MSP}

\bibliography{References}

@ARTICLE{Zou2021-on,
  title     = "Influence of flexoelectric effects on domain switching in
               ferroelectric films",
  author    = "Zou, M J and Tang, Y L and Feng, Y P and Geng, W R and Ma, X L
               and Zhu, Y L",
  journal   = "J. Appl. Phys.",
  publisher = "AIP Publishing",
  volume    =  129,
  number    =  18,
  pages     = "184103",
  month     =  may,
  year      =  2021,
  language  = "en"
}

@ARTICLE{Liang2024-uh,
  title     = "Advancements of flexoelectric materials and their
               implementations in flexoelectric devices",
  author    = "Liang, Xu and Dong, Huiting and Wang, Yifan and Ma, Qianqian and
               Shang, Hongxing and Hu, Shuling and Shen, Shengping",
  journal   = "Adv. Funct. Mater.",
  publisher = "Wiley",
  volume    =  34,
  number    =  51,
  month     =  dec,
  year      =  2024,
  copyright = "http://onlinelibrary.wiley.com/termsAndConditions\#vor",
  language  = "en"
}

@ARTICLE{Gu2015-uq,
  title     = "Nanoscale mechanical switching of ferroelectric polarization via
               flexoelectricity",
  author    = "Gu, Yijia and Hong, Zijian and Britson, Jason and Chen,
               Long-Qing",
  journal   = "Appl. Phys. Lett.",
  publisher = "AIP Publishing",
  volume    =  106,
  number    =  2,
  pages     = "022904",
  month     =  jan,
  year      =  2015,
  language  = "en"
}

@ARTICLE{Cao2017-nz,
  title     = "Pressure-induced switching in ferroelectrics: Phase-field
               modeling, electrochemistry, flexoelectric effect, and bulk
               vacancy dynamics",
  author    = "Cao, Ye and Morozovska, Anna and Kalinin, Sergei V",
  journal   = "Phys. Rev. B.",
  publisher = "American Physical Society (APS)",
  volume    =  96,
  number    =  18,
  month     =  nov,
  year      =  2017,
  copyright = "https://link.aps.org/licenses/aps-default-license"
}

@ARTICLE{Park2018-ow,
  title     = "Selective control of multiple ferroelectric switching pathways
               using a trailing flexoelectric field",
  author    = "Park, Sung Min and Wang, Bo and Das, Saikat and Chae, Seung Chul
               and Chung, Jin-Seok and Yoon, Jong-Gul and Chen, Long-Qing and
               Yang, Sang Mo and Noh, Tae Won",
  journal   = "Nat. Nanotechnol.",
  publisher = "Springer Science and Business Media LLC",
  volume    =  13,
  number    =  5,
  pages     = "366--370",
  month     =  may,
  year      =  2018,
  language  = "en"
}

@ARTICLE{Jia2025-il,
  title     = "Flexoelectric effect in thin films: Theory and applications",
  author    = "Jia, Xiaotong and Guo, Rui and Chen, Jingsheng and Yan, Xiaobing",
  journal   = "Adv. Funct. Mater.",
  publisher = "Wiley",
  volume    =  35,
  number    =  2,
  month     =  jan,
  year      =  2025,
  copyright = "http://onlinelibrary.wiley.com/termsAndConditions\#vor",
  language  = "en"
}

@ARTICLE{Harper1957-ug,
  title     = "The generation of static charge",
  author    = "Harper, W R",
  journal   = "Adv. Phys.",
  publisher = "Informa UK Limited",
  volume    =  6,
  number    =  24,
  pages     = "365--417",
  month     =  oct,
  year      =  1957,
  language  = "en"
}

@ARTICLE{Harper1970-hu,
  title     = "Triboelectrification",
  author    = "Harper, W R",
  journal   = "Phys. Educ.",
  publisher = "IOP Publishing",
  volume    =  5,
  number    =  2,
  pages     = "87--93",
  month     =  mar,
  year      =  1970
}

@ARTICLE{Stefani2021-hh,
  title     = "Mechanical reading of ferroelectric polarization",
  author    = "Stefani, Christina and Langenberg, Eric and Cordero-Edwards,
               Kumara and Schlom, Darrell G and Catalan, Gustau and Domingo,
               Neus",
  journal   = "J. Appl. Phys.",
  publisher = "AIP Publishing",
  volume    =  130,
  number    =  7,
  pages     = "074103",
  month     =  aug,
  year      =  2021,
  language  = "en"
}

@ARTICLE{Lu2012-hn,
  title     = "Mechanical writing of ferroelectric polarization",
  author    = "Lu, H and Bark, C-W and Esque de los Ojos, D and Alcala, J and
               Eom, C B and Catalan, G and Gruverman, A",
  journal   = "Science",
  publisher = "American Association for the Advancement of Science (AAAS)",
  volume    =  336,
  number    =  6077,
  pages     = "59--61",
  month     =  apr,
  year      =  2012,
  language  = "en"
}

@ARTICLE{Zhou2013-jc,
  title     = "In situ quantitative study of nanoscale triboelectrification and
               patterning",
  author    = "Zhou, Yu Sheng and Liu, Ying and Zhu, Guang and Lin, Zong-Hong
               and Pan, Caofeng and Jing, Qingshen and Wang, Zhong Lin",
  journal   = "Nano Lett.",
  publisher = "American Chemical Society (ACS)",
  volume    =  13,
  number    =  6,
  pages     = "2771--2776",
  month     =  jun,
  year      =  2013,
  language  = "en"
}

@ARTICLE{Cho2024-lp,
  title     = "Switchable tribology of ferroelectrics",
  author    = "Cho, Seongwoo and Gaponenko, Iaroslav and Cordero-Edwards,
               Kumara and Barcel{\'o}-Mercader, Jordi and Arias, Irene and Kim,
               Daeho and Lichtensteiger, C{\'e}line and Yeom, Jiwon and Musy,
               Lo{\"\i}c and Kim, Hyunji and Han, Seung Min and Catalan, Gustau
               and Paruch, Patrycja and Hong, Seungbum",
  journal   = "Nat. Commun.",
  publisher = "Springer Science and Business Media LLC",
  volume    =  15,
  number    =  1,
  pages     = "387",
  month     =  jan,
  year      =  2024,
  copyright = "https://creativecommons.org/licenses/by/4.0",
  language  = "en"
}

@ARTICLE{Heron2014-gu,
  title     = "Deterministic switching of ferromagnetism at room temperature
               using an electric field",
  author    = "Heron, J T and Bosse, J L and He, Q and Gao, Y and Trassin, M
               and Ye, L and Clarkson, J D and Wang, C and Liu, Jian and
               Salahuddin, S and Ralph, D C and Schlom, D G and I{\~n}iguez, J
               and Huey, B D and Ramesh, R",
  journal   = "Nature",
  publisher = "Springer Science and Business Media LLC",
  volume    =  516,
  number    =  7531,
  pages     = "370--373",
  month     =  dec,
  year      =  2014,
  language  = "en"
}

@article{zeches2009strain,
  title={A strain-driven morphotropic phase boundary in BiFeO3},
  author={Zeches, RJ and Rossell, MD and Zhang, JX and Hatt, AJ and He, Q and Yang, C-H and Kumar, A and Wang, CH and Melville, A and Adamo, C and others},
  journal={Science},
  volume={326},
  number={5955},
  pages={977--980},
  year={2009},
  publisher={American Association for the Advancement of Science}
}

@article{balke2012enhanced,
  title={Enhanced electric conductivity at ferroelectric vortex cores in BiFeO3},
  author={Balke, Nina and Winchester, Benjamin and Ren, Wei and Chu, Ying Hao and Morozovska, Anna N and Eliseev, Eugene A and Huijben, Mark and Vasudevan, Rama K and Maksymovych, Petro and Britson, Jason and others},
  journal={Nature Phys.},
  volume={8},
  number={1},
  pages={81--88},
  year={2012},
  publisher={Nature Publishing Group UK London}
}

@article{tagantsev2013origin,
  title={The origin of antiferroelectricity in PbZrO3},
  author={Tagantsev, Alexander K and Vaideeswaran, K and Vakhrushev, Sergey B and Filimonov, AV and Burkovsky, RG and Shaganov, A and Andronikova, D and Rudskoy, AI and Baron, AQR and Uchiyama, H and others},
  journal={Nature Commun.},
  volume={4},
  number={1},
  pages={2229},
  year={2013},
  publisher={Nature Publishing Group UK London}
}

@article{damjanovic1998ferroelectric,
  title={Ferroelectric, dielectric and piezoelectric properties of ferroelectric thin films and ceramics},
  author={Damjanovic, Dragan},
  journal={Reports on Progress in Physics},
  volume={61},
  number={9},
  pages={1267},
  year={1998},
  publisher={IOP Publishing}
}

@article{spaldin2005renaissance,
  title={The renaissance of magnetoelectric multiferroics},
  author={Spaldin, Nicola A and Fiebig, Manfred},
  journal={Science},
  volume={309},
  number={5733},
  pages={391--392},
  year={2005},
  publisher={American Association for the Advancement of Science}
}

@article{eerenstein2006multiferroic,
  title={Multiferroic and magnetoelectric materials},
  author={Eerenstein, Wilma and Mathur, ND and Scott, James F},
  journal={Nature},
  volume={442},
  number={7104},
  pages={759--765},
  year={2006},
  publisher={Nature Publishing Group UK London}
}

@article{ramesh2007multiferroics,
  title={Multiferroics: progress and prospects in thin films},
  author={Ramesh, Ramaroorthy and Spaldin, Nicola A},
  journal={Nature Materials},
  volume={6},
  number={1},
  pages={21--29},
  year={2007},
  publisher={Nature Publishing Group UK London}
}

@article{wang2003epitaxial,
  title={Epitaxial BiFeO3 multiferroic thin film heterostructures},
  author={Wang, JBNJ and Neaton, JB and Zheng, H and Nagarajan, V and Ogale, SB and Liu, B and Viehland, D and Vaithyanathan, V and Schlom, DG and Waghmare, UV and others},
  journal={Science},
  volume={299},
  number={5613},
  pages={1719--1722},
  year={2003},
  publisher={American Association for the Advancement of Science}
}

@article{catalan2009physics,
  title={Physics and applications of bismuth ferrite},
  author={Catalan, Gustau and Scott, James F},
  journal={Advanced Materials},
  volume={21},
  number={24},
  pages={2463--2485},
  year={2009},
  publisher={Wiley Online Library}
}

@article{sosnowska1982spiral,
  title={Spiral magnetic ordering in bismuth ferrite},
  author={Sosnowska, Izabela and Neumaier, T Peterlin and Steichele, E},
  journal={Journal of Physics C: Solid State Physics},
  volume={15},
  number={23},
  pages={4835},
  year={1982},
  publisher={IOP Publishing}
}

@article{kubel1990structure,
  title={Structure of a ferroelectric and ferroelastic monodomain crystal of the perovskite BiFeO3},
  author={Kubel, Frank and Schmid, Hans},
  journal={Structural Science},
  volume={46},
  number={6},
  pages={698--702},
  year={1990},
  publisher={International Union of Crystallography}
}

@article{wang2025manipulation,
  title={Manipulation of Ferroic Orders via Continuous Biaxial Strain Engineering in Multiferroic Bismuth Ferrite},
  author={Wang, Jiesu and Xu, Shuai and Meyer, Sebastian and Wu, Shiyao and Bandyopadhyay, Subhadeep and He, Xu and Miao, Qiyuan and Huang, Sisi and Li, Pengzhan and Zhao, Kun and others},
  journal={Advanced Science},
  volume={12},
  number={19},
  pages={2417165},
  year={2025},
  publisher={Wiley Online Library}
}

@article{gruverman2019piezoresponse,
  title={Piezoresponse force microscopy and nanoferroic phenomena},
  author={Gruverman, Alexei and Alexe, Marin and Meier, Dennis},
  journal={Nature Commun.},
  volume={10},
  number={1},
  pages={1661},
  year={2019},
  publisher={Nature Publishing Group UK London}
}


\end{document}